\documentclass[conference]{IEEEtran}
\IEEEoverridecommandlockouts
\usepackage{graphicx}
\usepackage{balance}  
\usepackage{xcolor}
\usepackage{amsfonts,amssymb,amsmath,bm}
\usepackage{graphicx}
\usepackage{makecell}
\usepackage{subfigure}
\usepackage{multirow}
\usepackage[ruled,vlined]{algorithm2e}
\usepackage{verbatim}

\usepackage[normalem]{ulem}
\useunder{\uline}{\ul}{}

\begin{document}


\title{CrowdTSC: Crowd-based Neural Networks for \\ Text Sentiment Classification}

\author{
    \IEEEauthorblockN{Keyu Yang$^{\dagger}$, Yunjun Gao$^{\dagger}$$^{\sharp}$, Lei Liang$^{\dagger}$, Song Bian$^{\dagger}$, Lu Chen$^{\ddagger}$, Baihua Zheng$^{*}$}
    \IEEEauthorblockA{
        \textit{{\large$^{\dagger}$}College of Computer Science, Zhejiang University, Hangzhou, China} \\
        \textit{{\large$^{\sharp}$}Alibaba--Zhejiang University Joint Institute of Frontier Technologies, Hangzhou, China} \\
        \textit{{\large$^{\ddagger}$}Department of Computer Science, Aalborg University, Denmark} \\
        \textit{{\large$^{*}$}School of Information Systems, Singapore Management University, Singapore, Singapore} \\
        \{kyyang, gaoyj, leiliang, songbian\}@zju.edu.cn ~~ luchen@cs.aau.dk ~~ bhzheng@smu.edu.sg
    }
}

\maketitle

\begin{abstract}
Sentiment classification is a fundamental task in content analysis. Although deep learning has demonstrated promising performance in text classification compared with shallow models, it is still not able to train a satisfying classifier for text sentiment. Human beings are more sophisticated than machine learning models in terms of understanding and capturing the emotional polarities of texts. In this paper, we leverage the power of human intelligence into text sentiment classification. We propose Crowd-based neural networks for Text Sentiment Classification (CrowdTSC for short). We design and post the questions on a crowdsourcing platform to collect the keywords in texts. Sampling and clustering are utilized to reduce the cost of crowdsourcing. Also, we present an attention-based neural network and a hybrid neural network, which incorporate the collected keywords as human being's guidance into deep neural networks. Extensive experiments on public datasets confirm that CrowdTSC outperforms state-of-the-art models, justifying the effectiveness of crowd-based keyword guidance.
\end{abstract}

\section{Introduction}
\label{sec:intro}

Sentiment classification \cite{NasukawaY03, DBLP:journals/widm/ZhangWL18} is a key content analysis task that has received much attention from both academia and industry. The goal is to assign emotional polarities (labels) to the specified text. Sentiment classification is essential for almost each and every human activity because it has a great impact on our decision making. With the explosive growth of social networks (e.g., Facebook, Twitter, Tumblr, and Sina Weibo), more and more individuals and organizations start using social media content to facilitate decision making. For example, merchants detect and analyze consumer insights by listening to consumers on social media; most consumers look at online reviews before they purchase; and governments employ social networks to understand public opinion on policy, etc.

Deep learning models \cite{DBLP:conf/nips/ZhangZL15, YogatamaDLB17, DBLP:conf/eacl/SchwenkBCL17, DBLP:conf/iclr/QiaoHNLD0Y018} have become the state-of-the-art solution for text classification. They learn to represent a text with an implicit vector and feed the vector into a \emph{softmax} function to calculate the probability of each class label. Recurrent Neural Network (RNN) and Convolutional Neural network (CNN) are two main kinds of neural networks used to represent the text. Deep neural networks learn the representation from all the words in the text without any guidance in advance. However, it is not a secret that certain words in a text are more important than the others in terms of text classification, and signals from some words provide an explicit indication about the class label. For instance, keyword \textit{happy} shows more \textit{positive} emotional polarity.


This suggests that if we can consider the keyword when training the deep neural networks for text sentiment classification, the accuracy might be improved. Here, we treat the keyword as the carrier of human intelligence, and it is well-known that human beings are more capable than machine learning (ML) algorithms in terms of capturing the emotional polarity of text. Recently, the development of crowdsourcing platforms, such as Amazon Mechanical Turk (AMT){\footnote{https://www.mturk.com}}, Figure Eight{\footnote{https://www.figure-eight.com}}, and Upwork{\footnote{https://www.upwork.com}}, offers a paradigm to collect intelligence from the crowd (i.e., thousands of ordinary workers). Nevertheless, how to efficiently collect and incorporate the intelligence into deep neural networks remains a big challenge.

In this paper, we investigate \emph{\underline{Crowd}-based neural networks for \underline{T}ext \underline{S}entiment \underline{C}lassification} (\emph{CrowdTSC} for short), which leverage the crowd wisdom to improve the performance of deep neural networks. There are two main challenges to be addressed. The first challenge is \textit{how to design cost-efficient crowd-based questions to capture the human being's guidance?} For sentiment classification, we could consult the crowd for every single text's sentiment classification exhaustively. Nonetheless, the brute-force approach is expensive, and lacks scalability as data are expected to arrive continuously. To this end, we introduce the concept of \textit{keyword}, which refers to the word that has a greater impact on the sentiment orientation than other words in the text, and only ask the crowd to identify keywords from the sampled texts. Then, we expand the keyword set based on clustering in the \textit{word embedding space}, by fully utilizing the fact that those words belonging to the similar semantic categories are proximal to each other in the embedding space~\cite{rohde2006}.

The second challenge is \textit{how to incorporate the collected keywords as human being's guidance into deep neural networks?} Deep neural networks are notorious for the un-interpretability. It is intractable to feed external intelligence guidance into deep neural networks. Towards this, we design two types of neural networks, namely, \emph{KA-RNN} and \emph{HDNN}, to embrace the collected keywords. KA-RNN is an attention-based RNN model whose \textit{loss function} has been redesigned to emphasize the keyword signals. HDNN is a hybrid deep neural network that combines a standard CNN (or RNN) with a Fully Connected Network (FCN) to integrate the information of original text with the keyword signals. To sum up, this paper makes the following four key contributions.

\begin{itemize}\setlength{\itemsep}{-\itemsep}
\item{} We present crowd-based neural networks for text sentiment classification. To our knowledge, it is the first attempt to utilize keywords collected from the crowd to improve the performance of deep learning for text classification.
\item{} We design a crowdsourcing framework to capture high-quality human being's guidance with a low monetary cost. In the framework, we utilize the proximity of similar semantic words in the embedding space, and then employ sampling and clustering techniques to reduce the cost, and meanwhile retain the performance.
\item{} We propose two models, i.e., KA-RNN and HDNN, to incorporate the collected keywords into deep neural networks. KA-RNN redesigns \textit{loss function} of attention-based RNN to emphasize the keyword signals, and HDNN builds a hybrid deep neural network that combines the standard CNN (or RNN) with FCN to enable the fusion of original text and keyword information.
\item{} We conduct extensive experiments to verify the effectiveness of our proposed CrowdTSC and the power of crowd-based keyword guidance compared with state-of-the-art models.
\end{itemize}

The rest of this paper is organized as follows. We first review related work in Section~\ref{sec:related}, and introduce the definition of sentiment classification and deep neural networks in Section~\ref{sec:pre}. We then elaborate the framework of CrowdTSC in Section~\ref{sec:overview}, present the cluster-based crowdsourcing in Section~\ref{sec:cluster}, and detail the customized attention-based RNN model and the hybrid deep neural network in Section~\ref{sec:attention} and Section~\ref{sec:hybrid}, respectively. Experimental evaluation is reported in Section~\ref{sec:experiment}. Finally, we conclude the paper in Section~\ref{sec:conclusion}.

\section{Related Work}
\label{sec:related}

\subsection{Crowdsourcing}

Nowadays, many important data management and analytics tasks can not be completely addressed by automated processes~\cite{DBLP:journals/tkde/LiWZF16}. Crowdsourcing is an effective technique to harness the capabilities of people (i.e., the crowd) to apply human computation for such tasks. The development of crowdsourcing platforms makes it an active research area in the data management community. Those crowdsourcing platforms allow computer scientists to integrate the power of human intelligence into their computational workflows. Take query processing as an example. Many crowd-based query processing systems have been implemented, such as CrowdDB~\cite{DBLP:journals/pvldb/FengFKKMRWX11},  Qurk~\cite{DBLP:conf/cidr/MarcusWMM11}, and Deco~\cite{DBLP:journals/pvldb/ParkPPGPW12}. They use optimization techniques to reduce the number of questions asked to crowd workers. In the field of image recognition, Welinder and Perona~\cite{DBLP:conf/cvpr/WelinderP10} propose a crowd-based algorithm to determine the ground truth for images from noisy annotations. For entity resolution, Vesdapunt et al.~\cite{DBLP:journals/pvldb/VesdapuntBD14} study the problem of completely resolving an entity graph using crowdsourcing; Wang et al.~\cite{DBLP:conf/sigmod/WangXL15} present ACD, a crowd-based algorithm for data deduplication, which achieves high accuracy at moderate costs of crowdsourcing. Besides, many studies~\cite{DBLP:conf/aaai/LinMW16, DBLP:journals/pvldb/MozafariSFJM14} apply active learning techniques to reduce the crowdsourcing cost for collecting and annotating data. Nevertheless, the above methods are application-dependent, and thus, they cannot be applied directly to tackle the text sentiment classification task.

\subsection{Text Sentiment Classification}

Traditional text classifiers are feature-based models, relying on hand-crafted features to perform the classification. They represent a text as a sparse vector, and feed it into the classifier. Cavnar et al.~\cite{cavnar1994n} propose an N-gram-based approach for text classification. Bag-of-words~\cite{Wallach06} is another efficient way to extract the features. Post and Bergsma~\cite{DBLP:conf/acl/PostB13} exploit more complex features such as~\emph{POS tagging} and \emph{dependency parsing} to improve the performance of text classification. Naive Bayes, maximum entropy classification, and support vector machines are popular classifiers~\cite{PangLV02}. Joulin et al.~\cite{DBLP:conf/eacl/GraveMJB17} show that simple linear models with a rank constraint and a fast loss approximation can achieve state-of-the-art performance. Nonetheless, feature-based models neglect the context of texts and hence cannot capture deep semantic information.



\begin{figure*}[ht]
    \centering
    \begin{minipage}{0.44\textwidth}
        \centering
        \includegraphics[width=0.9\textwidth]{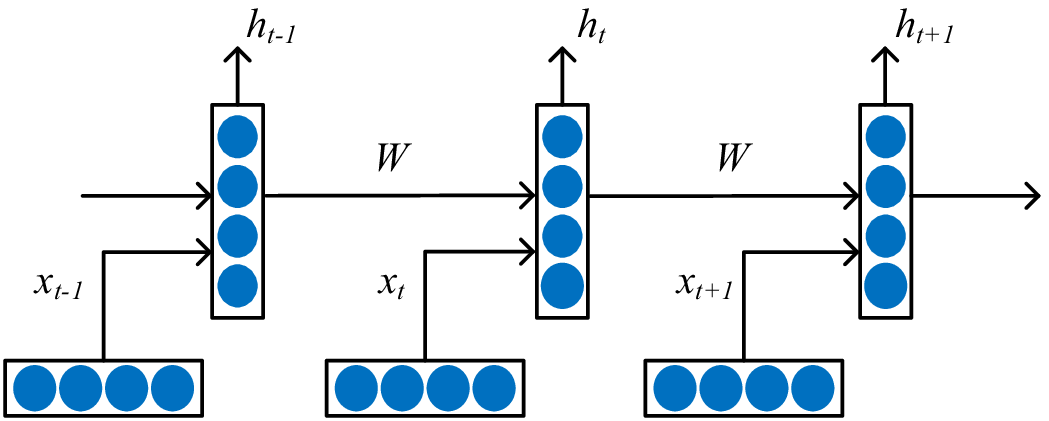}
        \caption{Illustration of RNN}
		\label{fig:RNN}
    \end{minipage}
    \begin{minipage}{0.46\textwidth}
        \centering
        \includegraphics[width=0.95\textwidth]{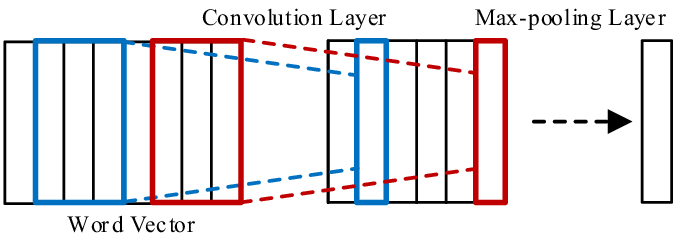}
		\caption{Illustration of CNN}
		\label{fig:CNN}
    \end{minipage}
\end{figure*}

To overcome such an issue, deep learning models have become popular for this task. They map a text to a dense vector. Most of the deep learning models are based on Convolutional Neural Network (CNN) or Recurrent Neural Network (RNN).
Zhang et al.~\cite{DBLP:conf/nips/ZhangZL15} present an empirical exploration of CNN for text classification.
Tang et al.~\cite{DBLP:conf/emnlp/TangQL15} leverage Long Short-Term Memory (LSTM)~\cite{DBLP:journals/neco/HochreiterS97} to model the relation of sentences.
Yang et al.~\cite{DBLP:conf/naacl/YangYDHSH16} propose a hierarchical attention network to better capture the important information of a document.
Conneau et al.~\cite{DBLP:conf/eacl/SchwenkBCL17} use very deep CNN in text classification, which achieves good performance. 
Yogatama et al.~\cite{YogatamaDLB17} present a discriminative LSTM model to place documents in the semantic space, such that embeddings of documents are close to embeddings of their respective labels.
Qiao et al.~\cite{DBLP:conf/iclr/QiaoHNLD0Y018} propose a method of learning and utilizing task-specific distributed representations of N-gram for text classification.
Wang et al.~\cite{DBLP:conf/sigir/WangHABQ19} propose a neural network model based on the context-aware attention mechanism for intent identification in E-mail conversations.
Islam et al. \cite{DBLP:conf/naacl/IslamM019} propose a multi-channel CNN architecture that effectively encodes different types of emotion indicators in social media posts for sentiment identification.
Recently, some researchers have attempted to combine CNN and RNN.
Wang et al.~\cite{DBLP:conf/acl/WangYLZ16} propose a regional CNN-LSTM model to compute valence-arousal ratings from texts for dimensional sentiment analysis.
Shi et al.~\cite{DBLP:conf/naacl/ShiYTJ16} replace convolution filters with deep LSTM.
Xiao and Cho ~\cite{DBLP:journals/corr/XiaoC16} utilize both convolution and recurrent layers to efficiently encode character inputs.
Wang and Wan~\cite{DBLP:conf/sigir/WangW18} propose a neural network with an abstract-based attention mechanism to address the sentiment analysis task of reviews for scholarly papers.

Although deep learning models achieve state-of-the-art performance for text sentiment classification, they do not make full use of the important signals carried by individual keywords. In this paper, we collect the keywords in the text via the crowdsourcing platform, and leverage the collective intelligence to guide deep neural networks to classify the text sentiment.



\section{Preliminaries}
\label{sec:pre}

\subsection{Sentiment Classification}

Sentiment classification (a.k.a. sentiment analysis or opinion mining) is a special task of text classification in Natural Language Processing (NLP) whose objective is to classify a text according to the sentimental polarities of opinions it contains. 
As an example, given a text about the movie review below:

\textit{``It's easily the best film I've seen this year, the story of the film is pretty great.''}

A machine learning model (classifier) could take this text as input, analyze the content, and assign the sentimental polarity (class), i.e., \textit{positive}, to this text.

\subsection{Deep Neural Networks}

\subsubsection{Recurrent Neural Network (RNN)}
RNN is a type of neural networks that conditions the model on all previous words in the corpus. Figure~\ref{fig:RNN} illustrates the RNN architecture where each rectangular box represents a hidden layer at a time step $t$. Each such layer holds a number of neurons, and performs a weighted sum operation on its inputs followed by a non-linear activation operation (such as $tanh()$, $sigmoid()$, and $ReLU()$). At each time step $t$, the output of the previous step $h_{t-1}$ and the next word embedding vector $x_t$ in the text will be input to the hidden layer to conduct the hidden representation $h_t$ in step $t$ as follows:
\begin{equation}\nonumber
h_{t}=\sigma\left(W^{(h h)} h_{t-1}+W^{(h x)} x_{t}\right)
\end{equation}
where $\sigma()$ is a non-linear activation function and both $W^{(h h)}$ and $W^{(h x)}$ are the weight matrices.



We can observe that the hidden representation of step $t$ depends upon all the previous input vectors. The $t$-th step state can be expressed by:
\begin{equation}\nonumber
 h_t = RNN(x_t, x_{t-1}, \cdots, x_1)
\end{equation}

The output of the hidden state in the last step could represent the text, and be the indicator of text classification.

Hidden states at each step depend on all the previous inputs, which sometimes neglect the key information and hurt the overall performance of the classifier~\cite{DBLP:journals/corr/0001KYS17}. Gating mechanisms have been developed to address the limitation of RNN, resulting in two prevailing RNN types, i.e., Long Short-Term Memory (LSTM)~\cite{DBLP:journals/neco/HochreiterS97} and Gated Recurrent Unit (GRU)~\cite{DBLP:conf/ssst/ChoMBB14}. Both LSTM and GRU can perform text classification, but we use GRU as the default RNN unit (detailed in Section~\ref{sec:attention}), because GRU is faster to train and more suitable for processing large-scale data.

\subsubsection{Convolutional Neural Network (CNN)}
Unlike RNN that models the whole sequence and captures the long-term dependencies, CNN is a class of neural networks that extracts local and position-invariant features. Figure~\ref{fig:CNN} depicts the CNN architecture. CNN takes word vectors, i.e., $d$-dimensional dense vectors, as input. It uses the convolution layer to represent learning from sliding $w$-grams. For an input sequence with $n$ word vectors, $x_1$, $x_2$, $\cdots$, $x_n$, let vector $c_{i} \in \mathbb{R}^{w d}$ be the concatenated embeddings of $w$ entries, $x_i$, $x_{i-1}$, $\cdots$, $x_{i-w+1}$, where $w$ is filter width and $w \le i \le n$. The convolution layer generates the representation $p_{i} \in \mathbb{R}^{d}$ for the $w$-gram $x_i$, $x_{i-1}$, $\cdots$, $x_{i-w+1}$ using the convolutional weights $W \in \mathbb{R}^{d \times w d}$:
\begin{equation}\nonumber
p_{i}=\sigma \left(W c_{i}+ b\right)
\end{equation}
where $\sigma()$ is a non-linear activation function and $b \in \mathbb{R}^{d}$ is the bias.

After the convolution layer, it uses max-pooling layer to extract the main information. For all $w$-gram representations $p_i$, a hidden representation $h_j$ is generated by max-pooling: $h_{j}=\max \left(p_{1, j}, p_{2, j}, \cdots\right), j=1, \cdots, d$. The hidden features could represent the text, and be the indicator of text classification.

\begin{figure*}[ht]
\centering
\includegraphics[width=0.94\textwidth]{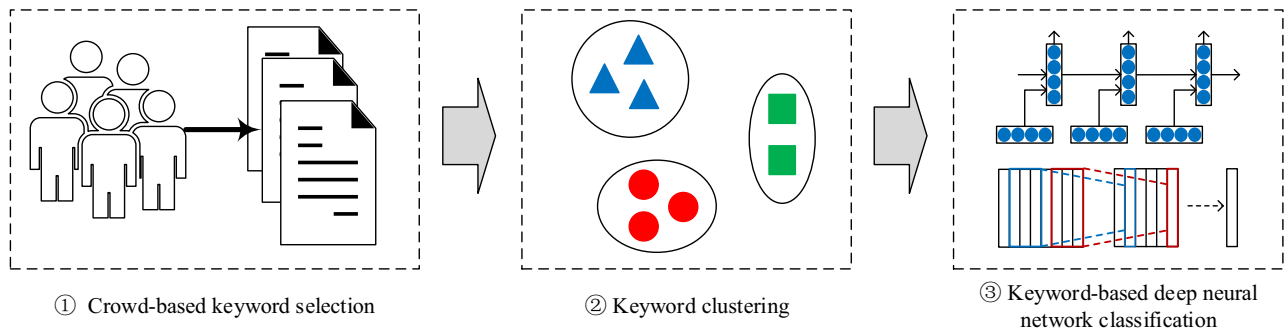}
\caption{Overview of CrowdTSC}
\label{fig:overview}
\end{figure*}

\section{Overview of CrowdTSC}
\label{sec:overview}

In this section, we overview our proposed CrowdTSC. As shown in Figure~\ref{fig:overview}, CrowdTSC consists of three stages, namely, \textit{crowd-based keyword selection}, \textit{keyword clustering}, and \textit{keyword-based deep neural network classification}.

In the first stage, we try to find the \textit{keywords} in the text through the human cognitive ability via the crowdsourcing platform. The keywords are the words in the text with a greater impact on the sentiment orientation than other words in the text. For the sake of low monetary cost, we sample the input text, and collect sampled keywords, instead of consulting the crowd workers for every single text's sentiment classification.

In the second stage, we expand the keyword set by using the clustering method. We understand the side-effect of sampling as it might reduce the positive impact of keywords on accuracy. In order to compromise the negative impact of sampling, we adopt a clustering approach to expand the keywords as a remedial action. The inspiration behind is that those words that belong to similar semantic categories are expected to be located in a neighboring area after being mapped into an embedding space~\cite{rohde2006}. Thus, we use a classic clustering method to capture word clusters in the embedding space, and expand the keyword set. Note that, the first and second stages will be detailed in Section~\ref{sec:cluster}.

In the last stage, the expanded keywords are utilized as collective intelligence from the crowd to guide deep neural networks. We design two different neural networks, i.e., KA-RNN and HDNN, to embrace the collected keywords. KA-RNN redesigns the \textit{loss function} for the attention-based RNN model to emphasize the keyword signals. HDNN is a hybrid deep neural network model that combines the standard CNN (or RNN) with FCN to fuse the original text information and keyword signals. We will detail these two models in Section~\ref{sec:attention} and Section~\ref{sec:hybrid}, respectively.
%

\section{Cluster-based crowdsourcing}
\label{sec:cluster}


\subsection{Sampling and Crowdsourcing}


As stated in Section~\ref{sec:intro}, it is impossible and unaffordable to ask the crowd to help in classifying \emph{each single} text in the corpus. Not to mention that many applications expect new texts to be continuously generated, while the brute-force approach lacks scalability. To tackle this issue, we propose a novel concept so-called \emph{keyword}, which refers to a word in a text that is informative and has a greater impact on the text's sentiment orientation than other words in the text. We regard keywords as the carriers of intelligence.

Accordingly, we consult crowd workers for the keywords in the text on the crowdsourcing platform. We adopt the sampling approach to sample only a small portion of the corpus to further reduce the monetary cost. Specifically, we post the crowdsourcing tasks on Amazon Mechanical Turk (AMT), and collect at least three keywords per sampled text from the crowd. As to be presented in Section \ref{sec:experiment}, our proposed method could achieve good performance even if only 0.1\% of the corpus is sampled.
%

\subsection{Clustering}

\begin{figure}[t]
\centering
\includegraphics[width=0.45\textwidth]{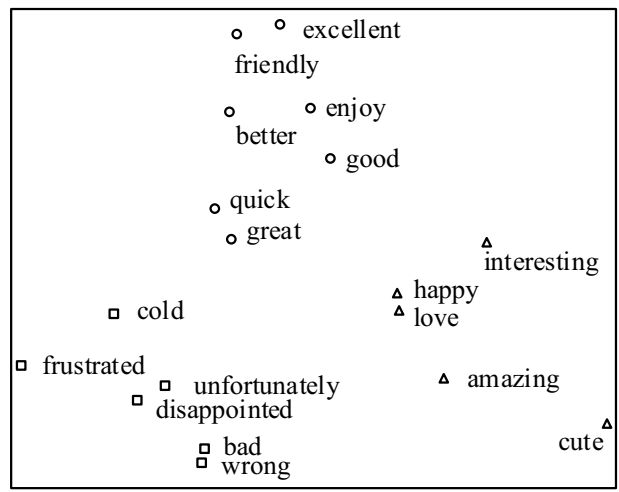}
\caption{A Case for the Word Embedding Space}
\label{fig:embedding}
\end{figure}

The main reason that the small sample size does not deteriorate the performance of our model is that we expand the small keywords set contributed by the crowd via clustering. The effectiveness of clustering is guaranteed by the fact that words sharing similar semantic meanings are expected to be close to each other in the embedding space~\cite{rohde2006}.

For illustration purposes, we adopt principal components analysis (PCA) to convert the embedding vector from high-dimensional space to 2-dimensional space, and visualize an embedding space example in Figure~\ref{fig:embedding}. It is observed that those words that have similar semantic meanings are close to each other. Take two words which have negative sentiment polarity,  i.e., \textit{unfortunately} and \textit{disappointed}, as an example. They are close and located on the lower left in Figure~\ref{fig:embedding}. Inspired by this observation, we adopt classic clustering methods~\cite{HanKP2011} to capture word clusters in the embedding space and to find the hidden keywords based on the clustering result with the help of the keywords contributed by the crowd. Next, we show how to identify the hidden keywords based on the clusters.

The main idea is to use the keywords collected from the crowd as the seeds to identify other keywords based on the clustering. To be more specific, we call keywords identified by the crowd as \emph{seed keywords}, and perform clustering based on seed keywords. We choose clusters that contain at least one seed keyword as \emph{keyword clusters}, and expand the keyword set based on those keyword clusters. Consider the embedding space depicted in Figure~\ref{fig:embedding} again. We could use the clustering method to group those words into three clusters, which are depicted as the circles in the upper, the squares in the lower left, and the triangles in the lower right, respectively. Assume that the crowd collect \textit{excellent}, \textit{cold}, and \textit{interesting} as seed keywords. All three clusters are labeled as keyword clusters, and we could include all the keywords into the expanded keyword set. That is to say, we expand the keyword set which contains 3 seed keywords to an expanded keyword set of 18 keywords.

\begin{algorithm}[t]
\caption{Keyword Expanding Algorithm (KEA)}
\label{algo:KEA}
\LinesNumbered
\DontPrintSemicolon
\KwIn{an embedding vector set $V$ for all the words, an embedding vector set $S$ for the seed keywords}
\KwOut{the expanded keyword embedding vector set $K$}
$K$ = $\emptyset$ \;
$\cup C_i$ = Cluster($V$)\;
\For{each $C_i$ in $\cup C_i$}{
    \If{$C_i \cap S \neq \emptyset$}{
        $K = K \cup C_i$ \;
    }
}
\Return the expanded keyword embedding vector set $K$ \;
\end{algorithm}

Based on these, we present \emph{Keyword Expanding Algorithm} (\emph{KEA}), with its pseudo-code listed in Algorithm~\ref{algo:KEA}. It takes as inputs an embedding vector set $V$ for all the words and an embedding vector set $S$ for seed keywords, and outputs the expanded keyword embedding vector set $K$. First, it initializes $K$ to an empty set (line 1). Then, it clusters the vector set $V$, with the resulting clusters preserved by $\cup C_i$ (line 2). Next, for each cluster $C_i \in \cup C_i$, KEA includes it into $K$ if it contains at least one seed keyword (lines 3-5). After evaluating all the clusters, KEA returns the expanded keyword set $K$ to complete the process (line 6).

Upon the completion of the first two stages of CrowdTSC, we generate an expanded keyword set, which is ready to be fed into deep neural networks to guide the sentiment classification. Next, we address the second challenge, which is how to incorporate human intelligence into the deep neural networks for text sentiment classification. In Section~\ref{sec:attention} and Section~\ref{sec:hybrid}, we propose two deep neural network models.

%

\section{Keyword-based RNN with Attention Mechanism}
\label{sec:attention}

\begin{figure}[t]
\centering
\includegraphics[width=0.475\textwidth]{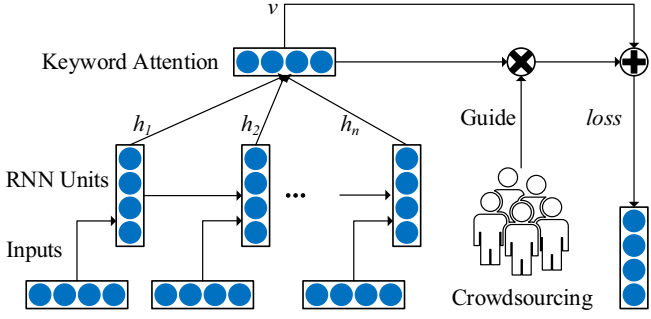}
\caption{The Architecture of KA-RNN}
\label{fig:KARNN}
\end{figure}

%
The first proposed deep learning model \emph{KA-RNN} is a type of RNN. It utilizes the attention mechanism~\cite{ChenHLXJ19, DBLP:conf/naacl/YangYDHSH16}, and takes into account the keywords collected by cluster-based crowdsourcing.
KA-RNN emphasizes the keyword signals by enabling keyword to play a greater impact on the attention weight. The structure of this model is shown in Figure \ref{fig:KARNN}. It contains three parts, namely, a \emph{word embedding input layer}, a \emph{standard RNN layer}, and an \emph{attention layer}. It uses the keywords that represent the human intelligence to guide the weight training of the attention layer, and combines the output in a fully connected layer to aggregate the loss. We describe the details in the following.

\subsection{Standard RNN}
\label{sec:RNN}
In the standard RNN layer, we use GRU \cite{DBLP:conf/ssst/ChoMBB14} to construct RNN. GRU employs a gating mechanism to capture potential long-term dependencies. The gating mechanism can control the flow of information, and mitigate the gradient vanishing problem. There are two types of gates in GRU, i.e., the \emph{reset gate} $r_t$ and the \emph{update gate} $z_t$. They control together how a hidden state is updated. At time step $t$, GRU computes $h_t$ as follows:
\begin{equation}\nonumber
h_{t}=\left(1-z_{t}\right) \odot h_{t-1}+z_{t} \odot \tilde{h}_{t}
\end{equation}
The computation is a linear combination of the previous state $h_{t-1}$ and the current new state $\tilde{h}_{t}$ that is derived from new input information, where $\odot$ is the element-wise multiplication. The gate $z_{t}$ decides how much past information shall be forgotten, and how much new information shall be considered. $z_{t}$ is computed as:
\begin{equation}\nonumber
z_{t}=\sigma\left(W_{z} x_{t}+U_{z} h_{t-1}+b_{z}\right)
\end{equation}
where $x_{t}$ is the input vector at time $t$, $W_{z}$ and $U_{z}$ are the weight parameters, and $b_{z}$ refers to the bias. The candidate new state $\tilde{h}_{t}$ is computed in a way similar as a traditional RNN:
\begin{equation}\nonumber
\tilde{h}_{t}=\sigma \left(W_{h} x_{t}+r_{t} \odot\left(U_{h} h_{t-1}\right)+b_{h}\right)
\end{equation}
where $r_{t}$ is the reset gate which controls how much the previous state contributes to the candidate new state, and $U_{h}$ is the weight matrix for $h_{t-1}$. Similar to the update gate, $r_{t}$ is computed as:
\begin{equation}\nonumber
r_{t}=\sigma\left(W_{r} x_{t}+U_{r} h_{t-1}+b_{r}\right)
\end{equation}
where $W_{r}$ and $U_{r}$ are the weight parameters, and $b_{r}$ is the bias.

\subsection{Attention Mechanism}

Each word in the text contributes differently to the representation of text. Standard RNN cannot differentiate the important words from the rest of the input text for text sentiment classification.
%
%
As a solution, we introduce the attention mechanism to extract important words, and describe how text sentiment classification can take advantage of keywords collected by cluster-based crowdsourcing.

Let $d$-dimensional vectors $h_{1}, h_{2}, \cdots, h_{n}$ denote the hidden representations produced by the standard RNN from the original input text, where $d$ is the size of the hidden layers and $n$ is the length of the input text. The attention mechanism will produce an attention weight vector $\alpha$ and a weighted hidden representation $v$. Specifically,
\begin{align*}
u_{i}&=\sigma \left(W_{w} h_{i}+b_{w}\right) \\
\alpha_{i}&=\frac{\exp \left(u_{i}^{\top} u_{w}\right)}{\sum_{i} \exp \left(u_{i}^{\top} u_{w}\right)} \\
v&=\sum_{i} \alpha_{i} h_{i}
\end{align*}
where $W_{w}$ and $b_{w}$ are the weight and bias parameters respectively.

%
%

We first feed word hidden representation $h_{i}$ using a non-linear active function to get $u_{i}$, and then, we measure the importance of the word as the similarity between $u_i$ and a word-level context vector $u_w$, and get a normalized importance weight $\alpha$ through a softmax function. The word context vector $u_w$ is randomly initialized and learned during the training process. After the attention weight vector $\alpha$ is produced, the vector $v$ is computed to summarize all the information of the input text. $v$ is considered as the feature representation of the input text. Then, a softmax function is to transform $v$ to a conditional probability distribution, in which $W_{t}$ and $b_{t}$ are the parameters of softmax function.
\begin{equation}\nonumber
p=softmax\left(W_{t} v+b_{t}\right)
\end{equation}

The attention weight vector $\alpha$ can be seen as a high-level representation of the query ``which is the informative word?''. When the word in $i$th state has a greater impact on the sentiment orientation than the word in $j$-th state, $\alpha_i$ would be larger than $\alpha_j$.

We could redesign the neural network and emphasize the keyword signals by using the attention weight vector $\alpha$. Here, the important design criterion is that the weight of the collected keyword should be larger than that of others. In other words, the keyword signals should be effectively amplified. In view of this, we design a new loss function for KA-RNN as follows:
\begin{equation}\nonumber
loss =-\sum_{d} \log p_{d j} - \lambda \sum_{d} m_{d}^T \alpha_d
\end{equation}
where $d$ refers to the input text, $j$ is the label of text $d$, $\lambda > 0$ is a penalty coefficient, and $m_d$ is a mask vector to indicate whether the current state in text $d$ is a collected keyword or not, i.e.,
\begin{equation}\nonumber
m_i=\left\{\begin{array}{ll}{1,} & {\text{if the } i\text{-th state inputs a collected keyword}} \\
{0,} & {\text {otherwise }}\end{array}\right.
\end{equation}
The loss function contains two parts: (i) the cross-entropy error between $p$ and the class label, and (ii) the regularization term for the attention weights. Since the goal of training is to minimize the loss function, the attention weight of the collected keyword, i.e., the keyword signal, tends to be amplified during the training process.

\section{Hybrid Deep Neural Network}
\label{sec:hybrid}

\begin{figure}[t]
\centering
\includegraphics[width=0.45\textwidth]{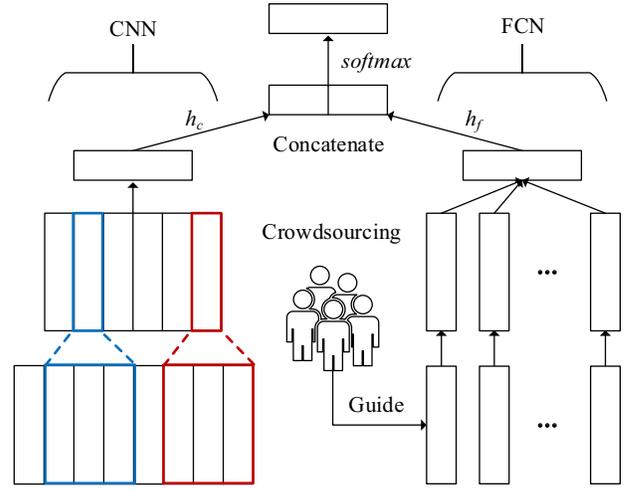}
\caption{The Architecture of HDNN}
\label{fig:HDNN}
\end{figure}

In this section, we propose our second deep neural network structure, which integrates the original text and the keyword information. It is a Hybrid Deep Neural Network (HDNN) that merges a standard CNN (or RNN) with an FCN. For the sake of brevity, we focus on the version of HDNN that is based on CNN and FCN with its architecture illustrated in Figure~\ref{fig:HDNN}. Note that CNN in HDNN could be replaced by RNN.

HDNN takes as inputs both the original text and the keywords collected by cluster-based crowdsourcing, and outputs the predicted class label. HDNN consists of two main components, a CNN and an FCN. It relies on the CNN to capture the hidden representation vector for the original text, and meanwhile, it invokes the FCN to obtain the hidden representation vector for the collected keywords. Then, it concatenates the two representation vectors to seamlessly fuse the original text information and human intelligence to effectively enhance the accuracy of the classification task. Finally, HDNN uses a softmax output layer to generate the probability for each class label. Next, we give the details of HDNN.
%

\subsection{Standard CNN}

Standard CNN is a key component of HDNN, as shown in the left of Figure~\ref{fig:HDNN}. It takes the original text $x$ as an input. Text $x$ contains $n$ words, with each corresponding to a $d$-dimensional word embedding vector. Thus, the input word embedding layer contains a feature map of $d \times n$ size. Next, it is the convolution layer which is used to extract the hidden representation from sliding $w$-grams. For the input word embedding with $n$ vectors, $x_1$, $x_2$, $\cdots$, $x_n$, let vector $c_{i} \in \mathbb{R}^{w d}$ be the concatenated embeddings of $w$ entries, $x_i$, $x_{i-1}$, $\cdots$, $x_{i-w+1}$, where $w$ is the filter width and $w < i < n$. The convolution layer generates the representation $p_{i} \in \mathbb{R}^{d}$ for the $w$-gram $x_i$, $x_{i-1}$, $\cdots$, $x_{i-w+1}$ using the convolutional weight $W \in \mathbb{R}^{d \times w d}$:
\begin{equation}\nonumber
p_{i}=ReLU \left(W c_{i}+ b\right)
\end{equation}
where $b \in \mathbb{R}^{d}$ represents the bias, and $ReLU$ is a type of active function:
\begin{equation}\nonumber
ReLU: f(z)= max(0,z)
\end{equation}

After the convolution layer, a max-pooling layer is used to extract the main information. For all $w$-gram representations $p_i$, a hidden feature $h_j$ is generated by max-pooling:
\begin{equation}\nonumber
h_{j}=\max \left(p_{1, j}, p_{2, j}, \cdots\right)(j=1, \cdots, d)
\end{equation}
The hidden feature $h_{j}$ could be seen as a high-level representation of the original text.

\subsection{Fully Connected Network}

The other main part of HDNN is an FCN, as depicted in the right of Figure~\ref{fig:HDNN}. After cluster-based crowdsourcing, we feed the keywords collected by cluster-based clustering into the FCN to capture the representation of the keywords and guide the sentiment classification. First, there is a fully connected layer that transforms $s$ $d$-dimensional keyword embedding vectors into $s$ hidden representations. For each keyword embedding $x_i$ ($i=1, 2, ..., s$), the fully connected layer computes the hidden representation $h_i$ as:
\begin{equation}\nonumber
h_{i}=ReLU \left(W_f x_{i}+ b\right)
\end{equation}
where $W_f \in \mathbb{R}^{d \times d}$ is the sharing weight matrix in the fully connected layer, and $b \in \mathbb{R}^{d}$ is the bias. After the fully connected layer, the max-pooling layer is again used to extract the main information. For each hidden representation $h_i$ ($i=1, 2, \cdots, s$), we capture the maximum value in $h_i$, and construct an $s$-dimensional vector $h_f$ to represent the keyword information. The vector $h_f$ is generated by:
\begin{equation}\nonumber
h_{fj}= \max \left(h_{1, j}, h_{2, j}, \cdots\right)(j=1, \cdots, d)
\end{equation}

\begin{table*}[ht]
\centering
\caption{Statistics of the Datasets Used in Our Experiments}\label{tab:dataset}
\begin{tabular}{lcccccc}
\hline
      Dataset     & Class Number & Train Samples  & Test Samples  & Average Length & Maximum Length & Vocabulary Size \\ \hline
\textit{AG's news}&       4      &     120,000    &     7,600     &      34        &      191       &     62,978      \\
\textit{Yelp.F}   &       5      &     650,000    &    50,000     &      148       &     1,169      &    268,271      \\
\textit{Yelp.P}   &       2      &     560,000    &    38,000     &      146       &     1,169      &    246,577      \\
\textit{Amazon.F} &       5      &    3,000,000   &    650,000    &      82        &      588       &  1,007,324      \\
\textit{Amazon.P} &       2      &    3,600,000   &    400,000    &      80        &      652       &  1,058,969      \\ \hline
\end{tabular}
\end{table*}

\subsection{Concatenation in HDNN}

We generate two different representations from two parts of HDNN to perform the text sentiment classification. One is the output of CNN, which captures the information of the original text. The other is the output of FCN, which extracts the key signals from the collected keywords. Next, we introduce how to fuse these two representations for the final classifier. We denote the output vector from the CNN as $h_c$, which is a $c$-dimensional vector, and the output vector from the FCN as $h_f$, which is an $f$-dimensional vector. Then, we use another fully connected layer to transform $h_c$ and $h_f$ into two $\frac{d}{2}$-dimensional vectors $h_{ct}$ and $h_{ft}$ respectively:
\begin{align*}
h_{ct}&= \left(W_{ct} h_{c}+ b\right) \\
h_{ft}&= \left(W_{ft} h_{f}+ b\right)
\end{align*}
where $W_{ct}\in \mathbb{R}^{c\times d/2}$, $W_{ft}\in \mathbb{R}^{f\times d/2}$, and bias $b \in \mathbb{R}^{d}$. The vector $h_{ct}$ can be seen as the hidden representation of the information from the original text, and the vector $h_{ft}$ carries the information extracted from the collected keywords.

We concatenate the two $d/2$-dimensional vectors into a $d$-dimen-sional vector $h_t$, which could serve as a high-level representation of the text classification with the guidance of human beings. Next, a softmax layer is followed to generate the probability distribution of predicted class labels:
\begin{equation}\nonumber
p=softmax\left(W_{t} h_t+b_{t}\right)
\end{equation}
where $W_{t}$ and $b_{t}$ are the parameters of the softmax function.

The model can be trained by backpropagation, in which the loss function is the cross-entropy loss. The loss function is computed as:
\begin{equation}\nonumber
loss =-\sum_{d} \log p_{d j}
\end{equation}
where $d$ is index of input text, and $j$ is the label of text $d$.

\section{Experimental Evaluation}
\label{sec:experiment}



\subsection{Experimental Settings}
\label{sec:expset}

We use five large-scale text datasets in~\cite{DBLP:conf/nips/ZhangZL15}, with their statistics listed in Table~\ref{tab:dataset}. \textit{AG's news} dataset consists of news obtained from AG's corpus. \textit{Yelp} dataset contains the reviews obtained from 2015 Yelp Dataset Challenge. \textit{Amazon} dataset is formed by reviews obtained from the Stanford Network Analysis Project (SNAP). Here, \textit{P} means the polarity prediction, while \textit{F} is full score prediction.

In our experiments, we sample 0.1\% of the datasets, and consult the crowd workers for the keywords that are important for sentiment classification in Amazon Mechanical Turk (AMT). Note that, \textit{AG's news} contains news articles from 4 different topics, and does not have sentimental polarities. We consult the crowd for the keywords that are high-related with the class labels, and treat \textit{AG's news} as a test for the generalization of CrowdTSC to normal text classification.
We utilize the 300D GloVe 42B vectors \cite{DBLP:conf/emnlp/PenningtonSM14} as our pre-trained word embeddings.
We implement three CrowdTSC models, viz., \emph{KA-RNN}, \emph{HDNN$_C$}, and \emph{HDNN$_R$}. KA-RNN is the model proposed in Section \ref{sec:attention}, HDNN$_C$ is one version of our proposed HDNN model discussed in Section~\ref{sec:hybrid}, and HDNN$_R$ the other version of HDNN model, which replaces the CNN architecture discussed in Section~\ref{sec:hybrid} with a standard RNN.
Our CrowdTSC models are implemented in Python 3.6 on Tensorflow 1.13.

\subsection{Performance Study}
\label{sec:exptech}

\subsubsection{The selection of clustering methods}

As discussed in Section~\ref{sec:cluster}, we employ the clustering method as an approach to expand the keyword set in order to guide the tuning of deep neural networks. This effectively reduces the number of crowd workers we have to approach, and thus, decreases the monetary cost of crowdsourcing.
%
%
In our experiments, we sample only 0.1\% of the original text dataset to consult the crowd workers for the keywords. After that, we need to select a clustering method to expand the keyword set. We evaluate the performance of four popular and classic clustering methods, using three datasets (viz., \textit{AG's news}, \textit{Yelp.F}, and \textit{Yelp.P}). Those four evaluated clustering methods could be grouped into two categories. One is centroid-based clustering methods, including $k$-means and Spectral~\cite{DBLP:journals/pami/ShiM00} clustering, and the other is density-based clustering methods, including DBSCAN~\cite{DBLP:conf/kdd/EsterKSX96} and mean-shift~\cite{DBLP:journals/pami/Cheng95} clustering.

%

\begin{table}
\centering
\caption{The Best Clustering Method}\label{tab:cluster}
\begin{tabular}{lcccc}
\hline
          & \multicolumn{4}{c}{\textit{AG's news}}                                             \\ \cline{2-5}
          & $k$-means      & Spectral        & DBSCAN         & Mean-shift    \\ \cline{2-5}
KA-RNN    & 91.13          & 91.24           & 91.27          & \textbf{92.04}\\
HDNN$_C$ & 91.86          & \textbf{92.04}  & 91.70          & 91.69          \\
HDNN$_R$ & 92.33          & 92.54           & \textbf{93.29} & 93.29          \\ \hline
          & \multicolumn{4}{c}{\textit{Yelp.F}}                                                \\ \cline{2-5}
          & $k$-means      & Spectral        & DBSCAN         & Mean-shift    \\ \cline{2-5}
KA-RNN    & 65.12          & \textbf{65.97}  & 65.03          & 65.08         \\
HDNN$_C$ & \textbf{64.32} & 63.26           & 63.33          & 63.33          \\
HDNN$_R$ & 65.00          & 64.38           & \textbf{66.97} & 65.01          \\ \hline
          & \multicolumn{4}{c}{\textit{Yelp.P}}                                                \\ \cline{2-5}
          & $k$-means      & Spectral        & DBSCAN         & Mean-shift    \\ \cline{2-5}
KA-RNN    & \textbf{96.47} & 96.18           & 96.22          & 96.12         \\
HDNN$_C$ & \textbf{95.83} & 95.76           & 95.82          & 95.77          \\
HDNN$_R$ & \textbf{96.58} & 96.33           & 96.44          & 95.41          \\ \hline

\end{tabular}
\end{table}

\begin{figure*}[t]
\centering
\includegraphics[width=0.85\textwidth]{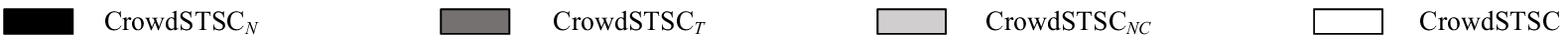}\\
  \subfigure[\textit{AG's news}]{
   \label{fig:cluster-AGnews}
   \includegraphics[width=0.31\textwidth]{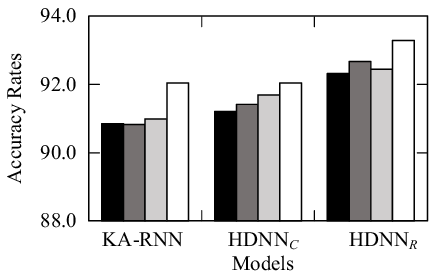}
  }
  \subfigure[\textit{Yelp.F}]{
   \label{fig:cluster-Yelp.F}
   \includegraphics[width=0.31\textwidth]{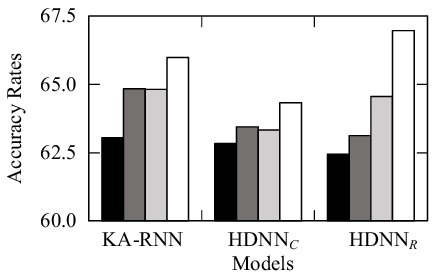}
  }
  \subfigure[\textit{Yelp.P}]{
   \label{fig:cluster-Yelp.P}
   \includegraphics[width=0.31\textwidth]{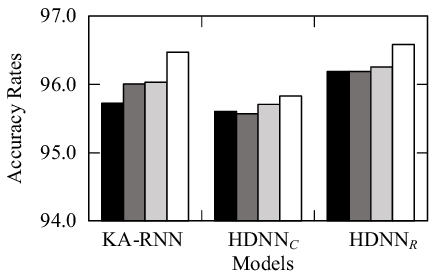}
  }
\caption{The Effect of Clustering-based Crowdsourcing}
\label{fig:cluster}
\end{figure*}

\begin{figure*}[t]
\centering
  \subfigure[\textit{AG's news}]{
   \label{fig:FCN-AGnews}
   \includegraphics[width=0.31\textwidth]{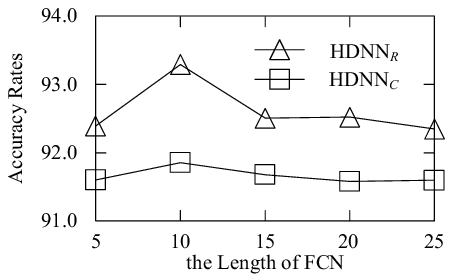}
  }
  \subfigure[\textit{Yelp.F}]{
   \label{fig:FCN-Yelp.F}
   \includegraphics[width=0.31\textwidth]{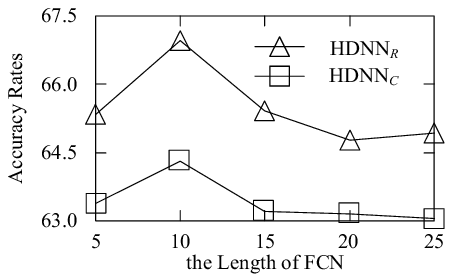}
  }
  \subfigure[\textit{Yelp.P}]{
   \label{fig:FCN-Yelp.P}
   \includegraphics[width=0.31\textwidth]{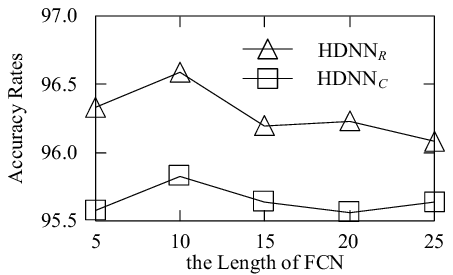}
  }
\caption{The Effect of FCN's Length in HDNN}
\label{fig:keywords}
\end{figure*}

We optimize the parameters for the clustering methods to provide competitive results. Table \ref{tab:cluster} lists the accuracy rates of our proposed models over different clustering methods using \textit{AG's news}, \textit{Yelp.F}, and \textit{Yelp.P} datasets. The results of the best clustering method are given in bold. Our experiments demonstrate that there is no single clustering method that can win out for all kinds of datasets. It once again verifies the ``no free lunch'' (NFL) theorem \cite{DBLP:journals/tec/DolpertM97} in the machine learning area. At the same time, we can observe that there is no remarkable difference between the results generated by different clustering methods. In other words, our cluster-based crowdsourcing could achieve satisfactory performance even by using a simple clustering method, e.g., $k$-means.

\subsubsection{The effect of clustering-based crowdsourcing}

In CrowdTSC, we employ the sampling technique to reduce the number of crowd workers we have to approach, and adopt a clustering method to expand the keywords. In order to verify the utility of keywords and the performance of clustering-based keyword expanding, we, in addition to CrowdTSC models proposed in this paper, implement three variants of CrowdTSC, denoted as \textit{CrowdTSC$_{N}$}, \textit{CrowdTSC$_{T}$}, and \textit{CrowdTSC$_{NC}$}, respectively. Here, we introduce the three additional variants of CrowdTSC.

\textit{CrowdTSC$_{N}$} refers to CrowdTSC \emph{without keywords}, i.e., we do not consider human intelligence for the task of text sentiment classification.

\textit{CrowdTSC$_{T}$} refers to CrowdTSC \emph{with keywords selected by a machine learning algorithm, i.e., TFIDF}, but not the human crowd. TFIDF, Term Frequency Inverse Document Frequency, is a numerical statistic that is intended to reflect how important a word is to a text\cite{DBLP:books/cu/LeskovecRU14}. We adopt TFIDF to pick $10$ most important keywords for every single text in the corpus without sampling.

\textit{CrowdTSC$_{NC}$} refers to CrowdTSC \emph{without clustering}. It still uses the crowd to help identify keywords in the sampled texts, but it does not adopt clustering algorithms to expand the keywords.

\begin{table*}
\center
\caption{Accuracy Rates on Five Datasets}\label{tab:result}
\begin{tabular}{llccccc}
\hline
Types                                 & Models       & \textit{AG's news}   & \textit{Yelp.F}      & \textit{Yelp.P}      & \textit{Amazon.F}    & \textit{Amazon.P}    \\ \hline
\multirow{4}{*}{Feature-based Models} & BoW~\cite{DBLP:conf/nips/ZhangZL15}    & 88.81                & 57.99          & 92.24                & 54.64          & 90.40                \\
                                     & BoW-TFIDF~\cite{DBLP:conf/nips/ZhangZL15}    & 89.64                & {\ul 59.86}          & 93.66                & {\ul 55.26}          & 91.00                \\
                                     & ngrams~\cite{DBLP:conf/nips/ZhangZL15}     & 92.04                & 56.26                & {\ul 95.64}          & 54.27                & {\ul 92.02}          \\
                                     & ngrams-TFIDF~\cite{DBLP:conf/nips/ZhangZL15} & {\ul 92.36}          & 54.80                & 95.44                & 52.44                & 91.54                \\ \hline
\multirow{7}{*}{Deep Learning Models} & char-CNN~\cite{DBLP:conf/nips/ZhangZL15}  & 90.49                & 62.05                & 95.12                & 59.57                & 95.07                \\
                                     &word-CNN~\cite{DBLP:conf/nips/ZhangZL15}   & 91.45                & 60.42                & 95.40                & 57.61                & 94.49                \\
                                     & char-CRNN~\cite{DBLP:journals/corr/XiaoC16}  & 91.36                & 61.82                & 94.49                & 59.23                & 94.13                \\
                                     & D-LSTM~\cite{YogatamaDLB17}     & 92.1                 & 59.6                 & 92.6                 & -                    & -                    \\
                                     & FastText~\cite{DBLP:conf/eacl/GraveMJB17}   & 92.5                 & 63.9                 & 95.7                 & 60.2                 & 94.6                 \\
                                     & VDCNN~\cite{DBLP:conf/eacl/SchwenkBCL17}      & 91.33                & 64.72                & 95.72 & {\ul \textbf{63.00}} & {\ul \textbf{95.72}} \\
                                     & Region.emb~\cite{DBLP:conf/iclr/QiaoHNLD0Y018} & {\ul 92.8}           & {\ul 64.9}           &  {\ul 96.4}                & 60.9                 & 95.3                 \\ \hline
\multirow{3}{*}{CrowdTSC}           & KA-RNN       & 92.04                & 65.97 & 96.47                & {\ul 60.32}          & {\ul 94.83}          \\
                                     & HDNN$_C$      & 92.04                & 64.32                & 95.83                & 56.44                & 93.21                \\
                                     & HDNN$_R$      & {\ul \textbf{93.29}} & {\ul \textbf{66.97}}                & {\ul \textbf{96.58}} & 58.94                & 93.92                \\ \hline
\end{tabular}
\end{table*}

The comparisons of the four versions of CrowdTSC are depicted in Figure~\ref{fig:cluster}, using \textit{AG's news}, \textit{Yelp.F}, and \textit{Yelp.P} datasets. In general, the models with keywords, either returned by TFIDF algorithm (i.e., CrowdTSC$_{T}$) or identified by human beings (i.e., CrowdTSC$_{NC}$ and CrowdTSC), outperform the version without keywords (i.e., CrowdTSC$_{N}$). This effectively demonstrates the positive impact of keywords on the accuracy of text sentiment classification. We also notice that CrowdTSC$_{T}$ performs even worse than CrowdTSC$_{N}$ in two out of nine cases, while CrowdTSC$_{NC}$ always outperforms CrowdTSC$_{N}$. It indicates that keywords, if selected wrongly, might have a negative impact on accuracy, while human beings are more capable than machines of locating the right keywords. When we compare the performance between CrowdTSC$_{NC}$ and CrowdTSC$_{T}$, we can observe that CrowdTSC$_{NC}$ achieves a higher accuracy rate than CrowdTSC$_{T}$ in five out of nine cases. Note that CrowdTSC$_{NC}$ only utilizes the keywords selected by human beings for $0.1\%$ of the texts in the corpus, while CrowdTSC$_{T}$ utilizes all the keywords returned by TFIDF for all the texts in the corpus. That is to say, the quality of keywords plays a much more important role in affecting accuracy than the number of keywords.

Consistent with our expectation, CrowdTSC performs the best in all nine cases. This signifies that when the keywords are able to reflect the sentiment of texts accurately, the number of keywords becomes important. The larger the number of properly selected keywords, the higher the accuracy rate of the classification task. Besides, CrowdTSC could effectively expand the proper keywords by using the clustering method.
%
%
%
In addition, it is observed that KA-RNN and HDNN$_R$ models that contain RNN units perform better than HDNN$_C$ model that does not have RNN units. It implies that RNN is more suitable to take advantage of collected keywords than CNN.

\subsubsection{The effect of FCN's length in HDNN}

Figure~\ref{fig:keywords} illustrates the accuracy rates of HDNN models w.r.t. the length of FCN varying from 5 to 25 on \textit{AG's news}, \textit{Yelp.F}, and \textit{Yelp.P} datasets. As discussed in Section \ref{sec:hybrid}, HDNN contains two main parts: CNN (or RNN) and FCN. The CNN (or RNN) part captures the information from the original text, while the FCN part extracts the collected keyword signals. The length of FCN is equivalent to the number of collected keywords that are input to HDNN. The first observation is that HDNN$_R$ exceeds HDNN$_C$ as expected. Besides, we can observe that the accuracy rates of HDNN (both HDNN$_C$ and HDNN$_R$) first ascend as the length is increased from a small value (e.g., 5), and then drop or stay stable as the length further grows. The optimal length of FCN is around 10. The reason is that the more the keywords are fed into the neural network, the more the important information it can learn. On the other hand, as the number becomes large, those collected keywords introduce noises to the model that could hurt the performance.
%

\subsection{Comparison with State-of-the-art Models}
\label{sec:expcomp}
\vspace{3.2mm}

Table~\ref{tab:result} lists the performance of our proposed models compared with the state-of-the-art models on five text classification datasets. For ease of discussion, we group all the models in three categories, including feature-based models, deep learning models, and CrowdTSC that refers to the models presented in this paper. They correspond to the three blocks in Table~\ref{tab:result}, respectively. The best results in each category are given in underscore, and the overall best records are given in bold.

The top block lists the performance of four feature-based models. These traditional models achieve a strong baseline accuracy rate in small datasets (including \textit{AG's news}, \textit{Yelp.F}, and \textit{Yelp.P}), but performs not well in large datasets (i.e., \textit{Amazon.F} and \textit{Amazon.P}).

The second block reports the performance of seven deep learning models. They achieve state-of-the-art performance using deep neural networks. It is observed that VDCNN performs best in large datasets, i.e., \textit{Amazon.F} and \textit{Amazon.P}. This is because VDCNN is a very deep CNN that uses up to 29 convolutional layers to extract the hidden representation for text classification. Therefore, VDCNN has also its own disadvantage, i.e., it is so deep that it is very sensitive to the parameter and difficult to tune.

The last block presents our proposed CrowdTSC models. The observation is that our presented models beat all the other models , except VDCNN, on all datasets including \textit{AG's news}. This is because we collect the keywords by clustering-based crowdsourcing, and embrace them as human guidance into the well-designed neural network architecture. This method is efficient to improve the accuracy of text sentiment classification, and could be extended to other general text classification. Besides, We win VDCNN in three datasets (i.e., \textit{AG's news}, \textit{Yelp.F}, and \textit{Yelp.P}), and lose in two datasets (i.e., \textit{Amazon.F} and \textit{Amazon.P}). Compared with the very deep and complex VDCNN model, our proposed CrowdTSC models are succinct, and thus, easy to tune.

%


\section{Conclusions}
\label{sec:conclusion}

In this paper, we propose Crowd-based neural networks for Text Sentiment Classification, i.e., CrowdTSC. To our knowledge, this is the first attempt to use keywords collected from the crowdsourcing platform to improve the performance of deep learning.
To reduce the monetary cost of hiring the crowd workers, we design a cluster-based crowdsourcing method to collect keywords in the given text datasets. Moreover, we develop two types of models to incorporate the collected keywords into deep neural networks, i.e., KA-RNN and HDNN. KA-RNN uses the attention mechanism, and constructs the loss function to emphasize keyword signals. HDNN combines the standard CNN (or RNN) with FCN to capture both the original text and the keyword information. Experimental results demonstrate that our proposed CrowdTSC models outperforms the state-of-the-art competitors, justifying the power of crowd-based human intelligence guidance.

\balance

\bibliographystyle{IEEEtran}
\bibliography{refer}

\end{document}